\documentclass[prd,twocolumn,showpacs,preprintnumbers,nofootinbib,amsmath,amssymb,floatfix]{revtex4}

\usepackage{graphicx}
\usepackage{dcolumn}
\usepackage{bm}
\usepackage{amsmath}

\newcommand{\beq}{\begin{eqnarray}}
\newcommand{\eeq}{\end{eqnarray}}

\newcommand{\real}{{\sf I}\kern-.12em{\sf R}}
\newcommand{\comp}{{\sf I}\kern-.50em{\sf C}}
\newcommand{\unity}{{\sf I}\kern-.54em{\sf 1}}

\def\spose#1{\hbox to 0pt{#1\hss}}
\def\ltapprox{\mathrel{\spose{\lower 3pt\hbox{$\mathchar"218$}}
 \raise 2.0pt\hbox{$\mathchar"13C$}}}

\begin{document}

\title{The phase diagram of QCD with four degenerate quarks}
\author{Paolo Cea}
\affiliation{
Dipartimento di Fisica dell'Universit\`a di Bari and INFN - Sezione di Bari, 
I-70126 Bari, Italy}
\email{paolo.cea@ba.infn.it}
\author{Leonardo Cosmai}
\affiliation{INFN - Sezione di Bari, I-70126 Bari, Italy}
\email{leonardo.cosmai@ba.infn.it}
\author{Massimo D'Elia}
\affiliation{Dipartimento di Fisica dell'Universit\`a di Genova 
and INFN - Sezione di Genova, I-16146 Genova, Italy}
\email{massimo.delia@ge.infn.it}
\author{Alessandro Papa}
\affiliation{Dipartimento di Fisica dell'Universit\`a della Calabria
and INFN - Gruppo collegato di Cosenza, 
I-87036 Arcavacata di Rende, Cosenza, Italy}
\email{papa@cs.infn.it}

\date{\today}

\begin{abstract}
We revisit the determination of the 
pseudo-critical line of QCD with four degenerate quarks at non-zero 
temperature and baryon density by the method of analytic continuation. 
We determine the pseudo-critical couplings at imaginary chemical 
potentials by high-statistics Monte Carlo simulations and reveal deviations 
from the simple quadratic dependence on the chemical potential visible in
earlier works on the same subject. Finally, we discuss the 
implications of our findings for the shape of the pseudo-critical line at real 
chemical potential, comparing different possible extrapolations.
\end{abstract}


\maketitle

\section{Introduction}
\label{introd}

The study of QCD at non-zero baryon density by numerical simulations
on a space-time lattice is plagued by the well-known sign problem:
the fermionic determinant is complex and the Monte Carlo sampling becomes
unfeasible.

One of the possibilities to circumvent this problem is to perform Monte
Carlo numerical simulations for imaginary values of the baryon 
chemical potential, where the fermionic determinant is real and 
the sign problem is absent, and to infer the behavior at real chemical 
potential by analytic continuation.

The idea of formulating a theory at imaginary $\mu$ was first suggested 
in Ref.~\cite{Alford:1998sd}, while the effectiveness of the method
of analytic continuation was pushed forward in Ref.~\cite{Lombardo:1999cz}.
Since then, the method has been extensively applied to  
QCD~\cite{muim,immu_dl,azcoiti,chen,defor06,Wu:2006su,sqgp,2im} and tested 
in QCD-like theories free of the sign problem~\cite{Hart:2000ef,giudice,
cea,cea1,conradi,Shinno:2009jw,cea2} and in spin models~\cite{potts3d,kt}. 

The state-of-the-art is the following: 
\begin{itemize}
\item the method is well-founded and works fine within the limitations posed 
by the presence of non-analyticities and by the periodicity of 
the theory with imaginary chemical potential~\cite{rw}; 
\item the analytic continuation of physical observables is improved if ratios 
of polynomials (or Pad\'e approximants~\cite{Lombardo:2005ks}) are used as 
interpolating functions at imaginary chemical potential~\cite{cea}; 
\item the analytic continuation of the (pseudo-)critical line on the
temperature -- chemical potential plane is well-justified, but careful
tests in two-color QCD~\cite{cea1,cea2} and in three-color QCD with finite 
isospin density~\cite{cea2} have evidenced some difficulties in its
application;
\item also some partial information about the nature of the phase transition as a function
of the chemical potential can be obtained by a careful study of the phase diagram in
the $T$ - Im($\mu$) plane~\cite{defor06,deph2,rwep}.
\end{itemize}

In particular, the numerical analyses in Refs.~\cite{cea1,cea2} 
have shown that, while there is no doubt that an analytic function exists 
which interpolates numerical data for the pseudo-critical couplings for both
imaginary and real $\mu$ across $\mu=0$, determining this function
by an interpolation of data at imaginary $\mu$ could be misleading.
Indeed, it was found that non-linear terms in the dependence of the 
pseudocritical coupling $\beta_c$ on $\mu^2$ in general cannot be neglected and
the prediction for the pseudo-critical couplings at real chemical potentials 
may be wrong if data at imaginary $\mu$ are fitted according to a linear 
dependence. Moreover, the coefficients of the linear and non-linear terms in 
$\mu^2$ in a Taylor expansion of $\beta_c(\mu^2)$ are all negative. That 
often implies subtle cancellations of non-linear terms at imaginary chemical 
potentials ($\mu^2 < 0$) in the region available for analytic
continuation (first Roberge-Weiss sector). The detection of such terms,
from simulations at $\mu^2 < 0$ only, may be difficult and requires an
extremely high accuracy.

In Refs.~\cite{cea1,cea2} it was realized that, though a polynomial of order 
$\mu^6$ seems to be sufficient in all explored cases in two-color QCD and in
three-color QCD at finite isospin density, an increased predictivity can be achieved 
by a fit with the linear term in $\mu^2$ fixed from data at small values
of $\mu^2$ only and kept ``constrained'' when data at larger values of 
$\mu^2$ are included. Moreover, the idea was proposed to parameterize the 
critical line directly in physical units in the $T,\mu$ plane (instead than 
in the $\beta,\mu$ plane), and some Ans\"atze were tested for such 
parameterization, which provided a very good description of the critical line 
with a reduced number of parameters and an increased predictivity.

The aim of this paper is to apply the experience acquired through the 
study of sign-problem-free theories to the determination of the 
pseudo-critical line  in three-color QCD with four degenerate quarks. With 
respect to the well-known work of Ref.~\cite{immu_dl}, we will take advantage 
of the use of a smaller lattice for collecting more determinations of 
pseudo-critical couplings at imaginary chemical potential with increased 
numerical accuracy and will apply on the new lattice data the fit strategies 
worked out in our previous studies of Refs.~\cite{cea1,cea2}.

\section{Numerical results}
\label{results}

In our numerical analysis, we consider SU(3) with $n_f=4$ degenerate 
standard staggered fermions of mass $am=0.05$ on a $12^3\times 4$ lattice. 
The algorithm adopted for Monte Carlo numerical simulations is the 
exact $\Phi$ algorithm described in Ref.~\cite{Gottlieb:1987mq}, properly 
modified for the inclusion of a finite chemical potential. 

\begin{figure}[htbp]
\includegraphics*[height=0.295\textheight,width=0.95\columnwidth]
{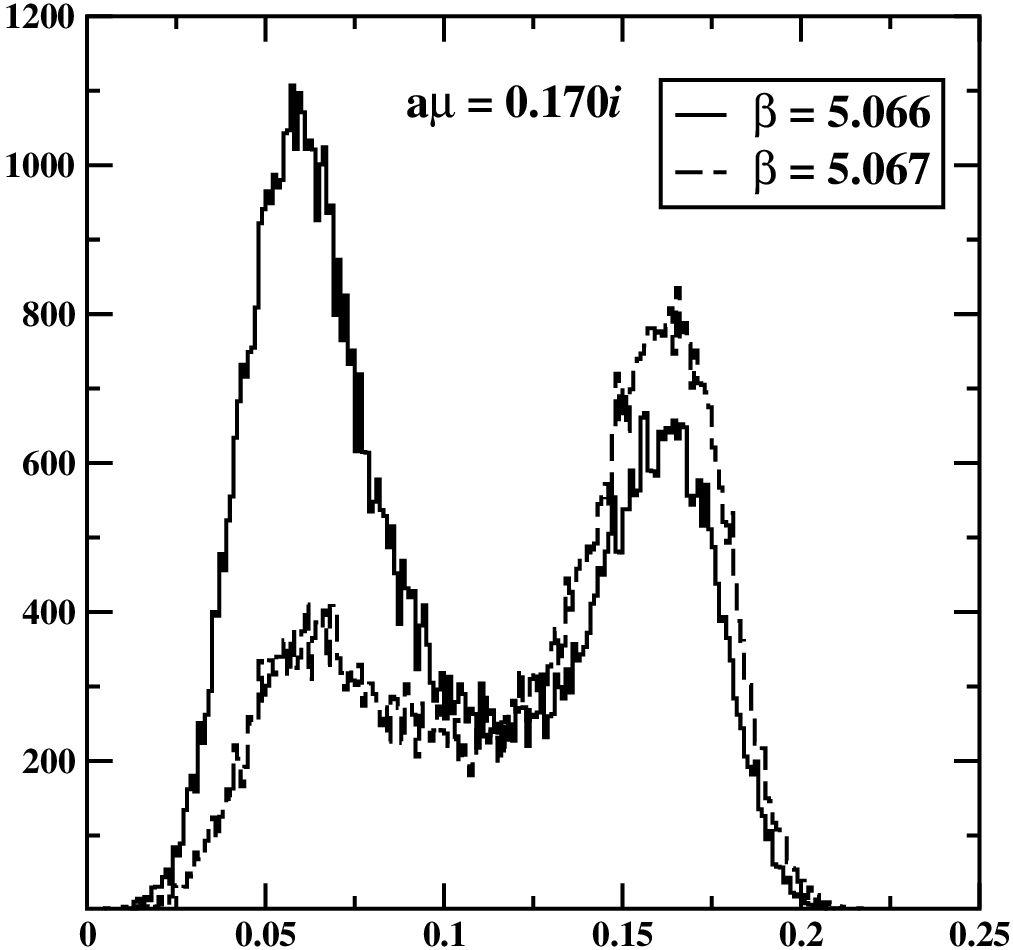}
\caption{Distribution of the real part of the Polyakov loop 
in SU(3) with $n_f=4$ on a 12$^3\times 4$ lattice with $am$=0.05 
at $a\mu=0.170 i$ and for two $\beta$ values around the transition.}
\label{su3_distribution}
\end{figure}

\begin{figure}[htbp]
\includegraphics*[height=0.295\textheight,width=0.95\columnwidth]
{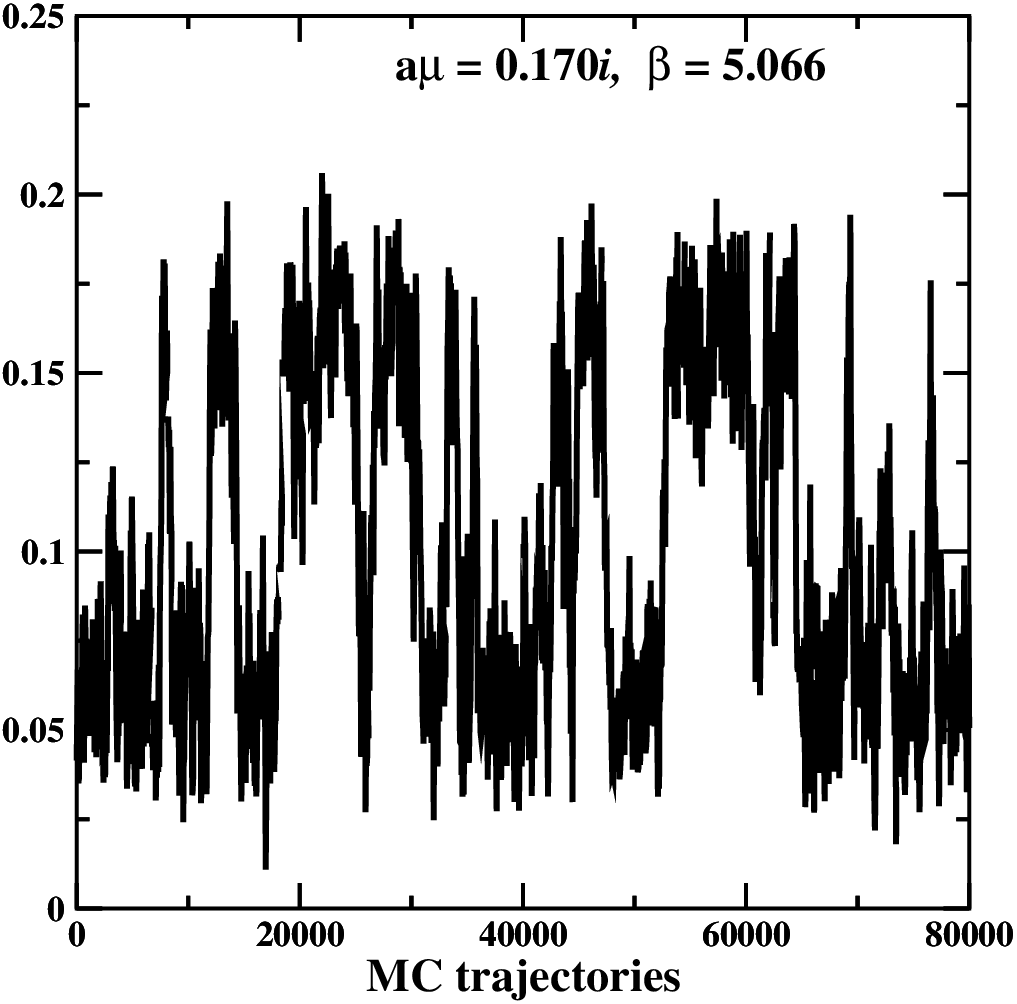}
\caption{Monte Carlo history of the real part of the Polyakov loop 
in SU(3) with $n_f=4$ on a 12$^3\times 4$ lattice with 
$am$=0.05 at $a\mu=0.170i$ and $\beta$=5.066.}
\label{su3_poly_history}
\end{figure}

\begin{figure}[htbp]
\includegraphics*[height=0.295\textheight,width=0.95\columnwidth]
{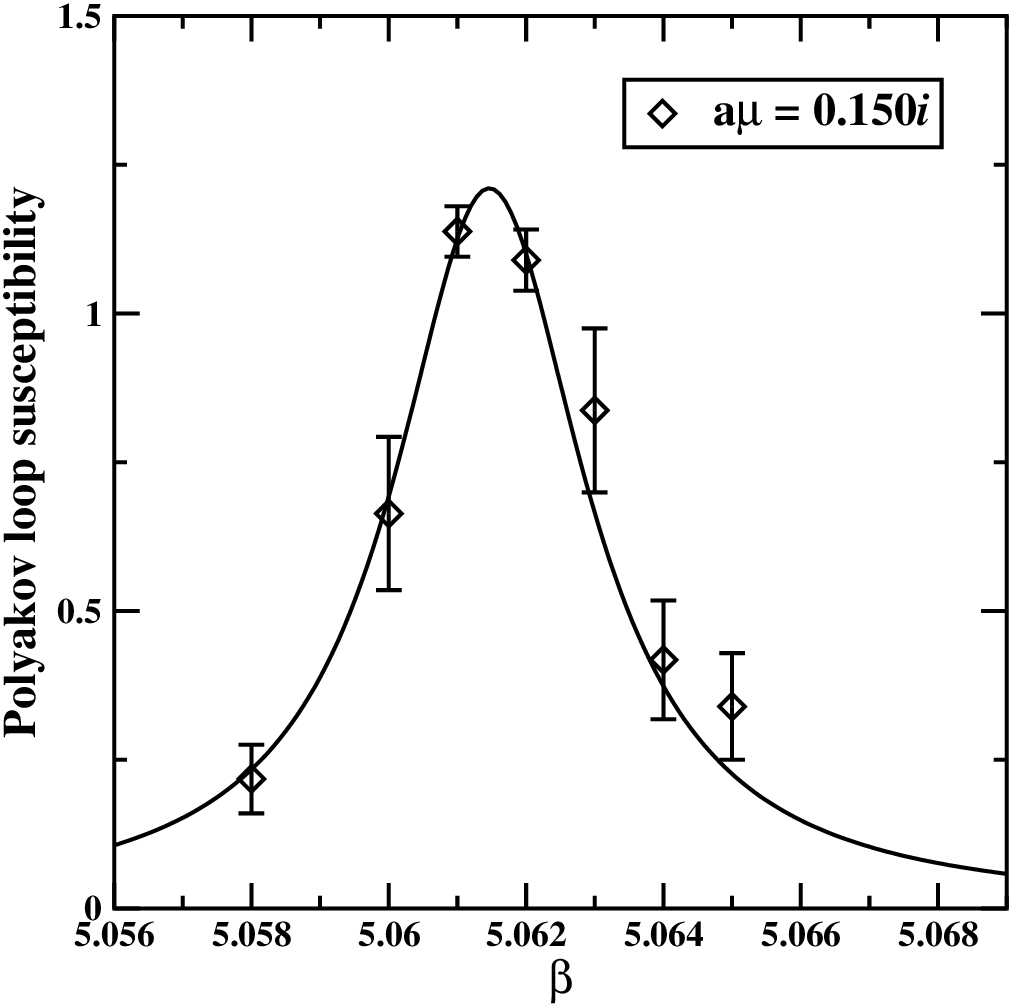}
\caption{Susceptibility of the (real part of the) Polyakov loop
{\it vs} $\beta$ in SU(3) with $n_f=4$ on a 12$^3\times 4$ lattice with 
$am$=0.05 and $a\mu = 0.150i$. The solid lines represent the Lorentzian 
interpolation.}
\label{su3_peak}
\end{figure}

\begin{figure}[htbp]
\includegraphics*[height=0.295\textheight,width=0.95\columnwidth]
{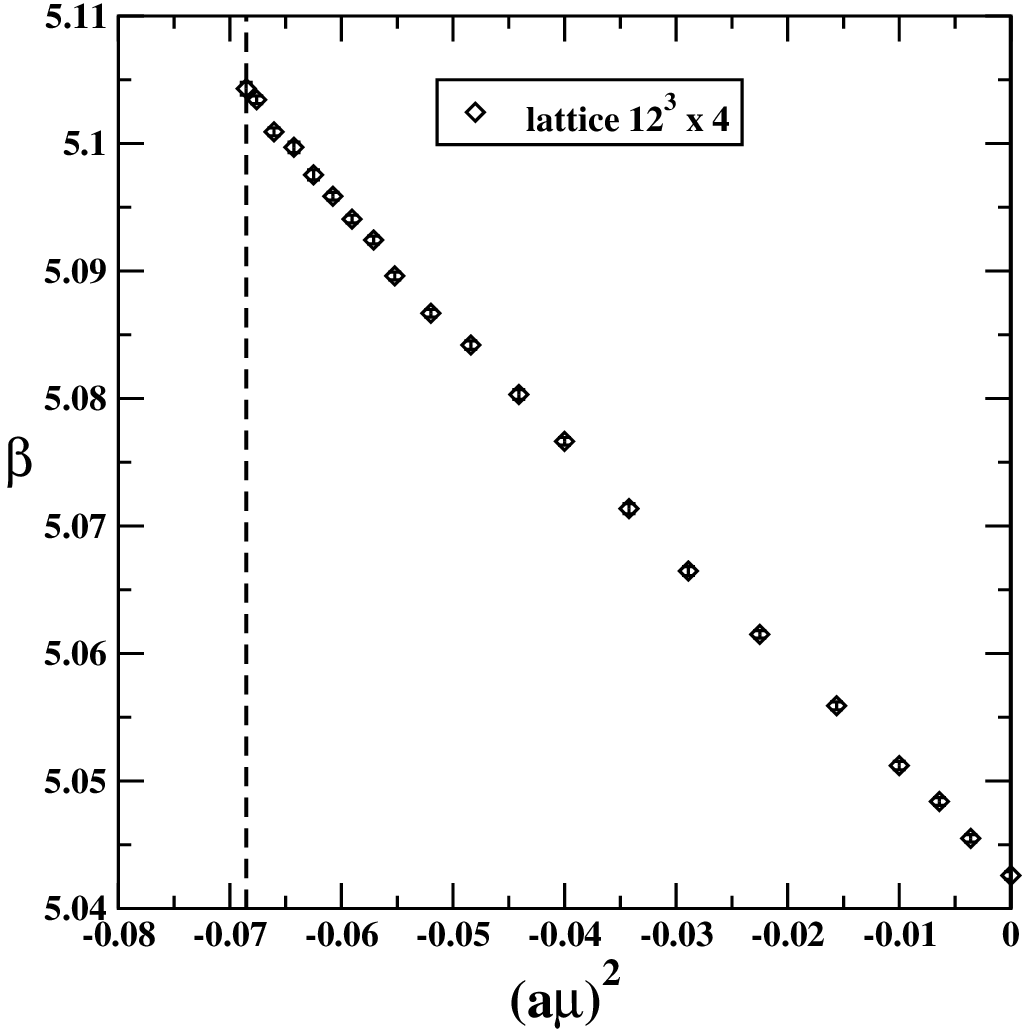}
\caption{Critical couplings obtained in SU(3) with $n_f=4$ 
on a 12$^3\times 4$ lattice with $am$=0.05. The dashed vertical line
indicates the boundary of the first RW sector, $a\, {\rm Im}(\mu)=\pi/12$.}
\label{su3_beta_crit}
\end{figure}

In this theory the critical line in the temperature -- chemical potential 
plane is a line of first order transitions, over all the range of 
$\mu^2$ values in the first Roberge-Weiss (RW) sector, $-(\pi/3)^2 
\leq (\mu/T)^2 \leq 0$. Tunneling between the different phases 
clearly emerges from the distribution on the thermal equilibrium ensemble of 
the values of observables like the (real part of) the Polyakov loop, the 
chiral condensate, the plaquette across the transition (see 
Fig.~\ref{su3_distribution} as an example of the typical two-peak 
structures). As a further evidence of the first order nature of the 
transition, we show in Fig.~\ref{su3_poly_history} the Monte Carlo run 
history of the (real part of) the Polyakov loop at $a\mu =0.170i$ 
and $\beta=5.066 \simeq \beta_c$, which exhibits tunneling events between 
the two phases every few thousands trajectories, on average. 
Typical statistics have been around 10k trajectories of 1 Molecular Dynamics
unit for each run, growing up to 100k trajectories for 4-5 $\beta$ values
around $\beta_c(\mu^2)$, for each $\mu^2$, in order to correctly sample
the critical behavior at the transition.
The critical $\beta(\mu^2)$ is determined as the value for which the 
susceptibility of the (real part of the) Polyakov loop exhibits 
a peak. To precisely localize the peak, a Lorentzian interpolation is used
(see Fig.~\ref{su3_peak}, for example).
In all cases, this kind of determination is compatible with that consisting
in estimating the point where the two peaks in the distribution of
the (real part of the) Polyakov loop have equal height, or in locating
the peak of the susceptibility by the Ferrenberg-Swendsen method. 
 We verified also
that the determinations do not change if the susceptibility of a 
different observable, such as the baryon number, is used. In 
Table~\ref{su3_data} we summarize our determinations of the critical couplings:
in a few cases we have repeated the determination also on a $16^3 \times 4$ 
lattice, where negligible corrections, within the reported errors, have been 
observed. The plot of the data for $\beta(\mu^2)$ versus $(a\mu)^2$ -- see
Fig.~\ref{su3_beta_crit} -- clearly shows that data do not line up along a
straight line in all the first RW sector, thus indicating that the curve
$\beta(\mu^2)$ cannot be parametrized by a polynomial of order $\mu^2$.
In fact, as we will see soon, at least a polynomial of order $\mu^6$ is
needed to get a fit with a reasonable $\chi^2$.

\begin{table}[htbp]
\setlength{\tabcolsep}{0.5pc}
\centering
\caption[]{Summary of the values of $\beta_c(\mu^2)$ for SU(3) with 
$n_f=4$ on the 12$^3\times 4$ lattice with fermionic mass $am$=0.05.}
\begin{tabular}{dd}
\hline
\hline
\multicolumn{1}{c}{\hspace{0.70cm}$a\,{\rm Im}(\mu)$} &
\multicolumn{1}{c}{\hspace{1cm}$\beta_c$} \\
\hline
0.     & 5.04259(30) \\
0.060  & 5.04550(30) \\
0.080  & 5.04839(30) \\
0.100  & 5.05121(33) \\
0.125  & 5.05590(31) \\
0.150  & 5.06150(30) \\
0.170  & 5.06647(35) \\
0.185  & 5.07136(40) \\
0.200  & 5.07664(30) \\
0.210  & 5.08031(38) \\
0.220  & 5.08419(33) \\
0.228  & 5.08668(30) \\
0.235  & 5.08961(30) \\
0.239  & 5.09243(30) \\
0.243  & 5.09407(30) \\
0.2465 & 5.09586(30) \\
0.250  & 5.09754(40) \\
0.2535 & 5.09970(42) \\
0.257  & 5.10092(31) \\
0.260  & 5.10343(30) \\
\multicolumn{1}{c}{\hspace{0.3cm}$\pi/12$} & 5.1043(5) \\
\hline
\hline
\end{tabular}
\label{su3_data}
\end{table}

\begin{figure*}[htbp]
\includegraphics*[height=0.295\textheight,width=0.95\columnwidth]
{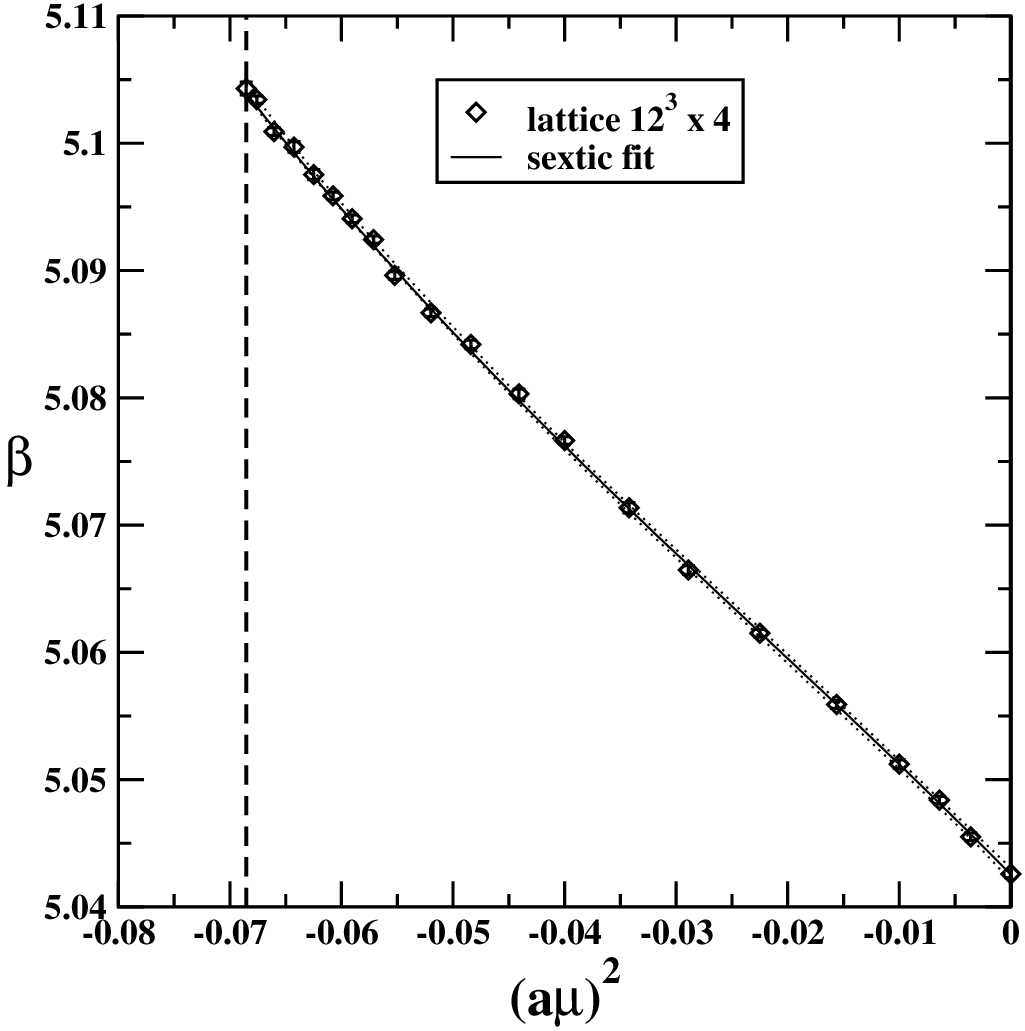}
\includegraphics*[height=0.295\textheight,width=0.95\columnwidth]
{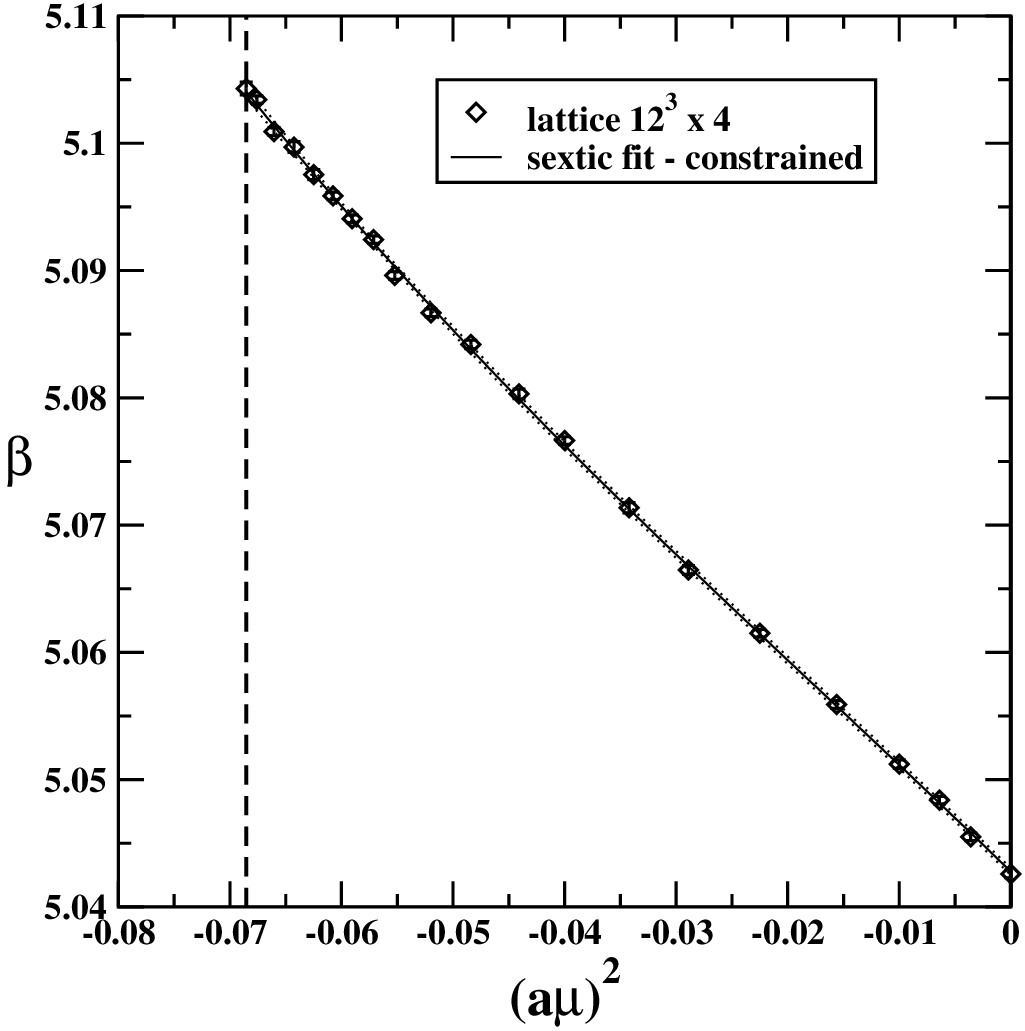}
\caption{Fits to the critical couplings: plain 6th order polynomial (left) 
and 6th-order polynomial with constrained coefficient of the quadratic term 
(right).}
\label{fig_fits}
\end{figure*}

We have tried several kind of interpolations of the critical couplings 
at $\mu^2 \leq 0$. 
At first, we have considered interpolations with polynomials up to
order $\mu^{6}$ (see Table~\ref{table_fits} for a summary of the
resulting fit parameters and their uncertainties as obtained with the 
MINUIT minimization code). We can see that data at $\mu^2 \leq 0$ are 
precise enough to be sensitive to terms beyond the order $\mu^2$; indeed,
a good $\chi^2$/d.o.f. is not achieved before including terms up to the 
order $\mu^6$. In Fig.~\ref{fig_fits}(left) we show how the fit with the
6th-order polynomial compares with data of $\beta_c(\mu^2)$; the dotted
lines around the fitting curve (solid line) delimit the 95\% CL band.

\begin{table*}[htbp]
\setlength{\tabcolsep}{0.13pc}
\centering
\caption[]{Parameters of the fits to the critical couplings 
in SU(3) with $n_f=4$ on a 12$^3\times 4$ lattice with fermionic 
mass $am$=0.05, according to the fit function
$\beta_c (\mu^2) = (a_0 + a_1 (a\mu)^2 + a_2 (a\mu)^4 + a_3 (a\mu)^6)
/(1 + a_4 (a\mu)^2 + a_5 (a\mu)^4)$. 
Blank columns stand for terms not included in the fit. The asterisk
denotes a constrained parameter. Fits are performed in the interval
$[(a\mu_{\rm min})^2,0]$; the last column gives the value of 
$(a\mu_{\rm min})^2$.}
\begin{tabular}{ddddddcc}
\hline
\hline
\multicolumn{1}{c}{\hspace{1cm}$a_0$} & \multicolumn{1}{c}{\hspace{1cm}$a_1$} &
\multicolumn{1}{c}{\hspace{1cm}$a_2$} & \multicolumn{1}{c}{\hspace{1cm}$a_3$} &
\multicolumn{1}{c}{\hspace{1cm}$a_4$} & \multicolumn{1}{c}{\hspace{1cm}$a_5$} & $\chi^2$/d.o.f. & $(a\mu_{\rm min})^2$ \\
\hline
5.04198(22) & -0.8839(48) &           &            &             &           & 6.63 & $-(\pi/12)^2$ \\
5.04256(24) & -0.8509(71) &           &            &             &           & 0.85 & $-0.235^2$ \\
5.04311(36) & -0.761(26)  & 1.77(36)  &            &             &           & 2.13 & $-(\pi/12)^2$ \\
5.04254(50) & -0.892(72)  & -3.1(2.4) & -46.(23.)  &             &           & 1.10 & $-(\pi/12)^2$ \\
5.04277(27) & -0.8509^*   & -1.70(55) & -34.0(8.2) &             &           & 1.20 & $-(\pi/12)^2$ \\
5.04284(28) & 55.799(14)  &           &            & 11.2266(27) & 1.741(29) & 1.13 & $-(\pi/12)^2$ \\
5.04276(27) & 58.196(12)  & -9.46(13) &            & 11.7044(24) &           & 1.09 & $-(\pi/12)^2$ \\
\hline
\hline
\end{tabular}
\label{table_fits}
\end{table*}

As in Ref.~\cite{cea2}, we performed a ``constrained'' fit: first,
the largest interval $[(a\mu)^2_{\rm min},0]$ was identified where data
could be interpolated by a polynomial in $\mu^2$, with a $\chi^2$/d.o.f
$\sim 1$. It turned out that 
$(a\mu)^2_{\rm min}=(0.235i)^2$, the quadratic coefficient being equal to 
$-0.8509$. Then, all available data were fitted by a 6th-order polynomial,
with the quadratic coefficient fixed at $-0.8509$ 
(see Table~\ref{table_fits} and Fig.~\ref{fig_fits}(right)). 

\begin{figure*}[htbp]
\includegraphics*[height=0.295\textheight,width=0.95\columnwidth]
{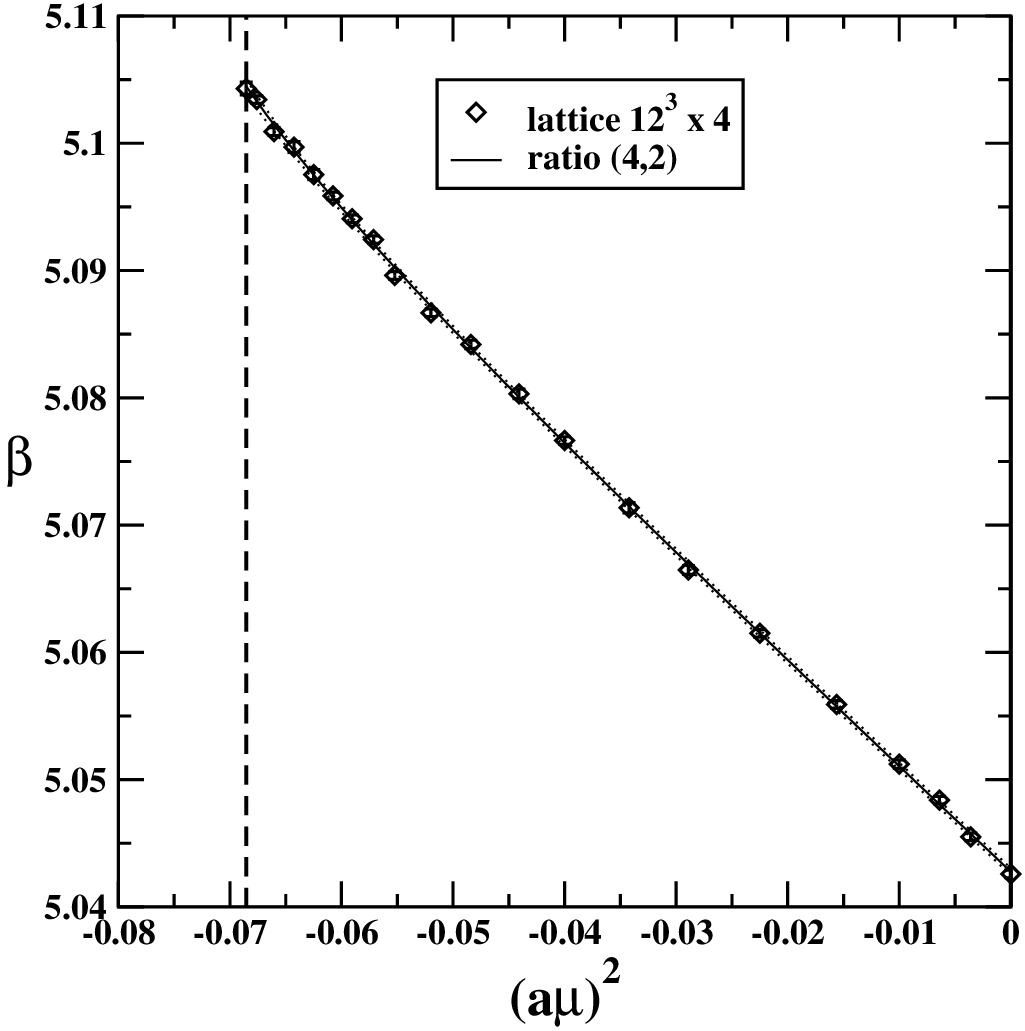}
\includegraphics*[height=0.295\textheight,width=0.95\columnwidth]
{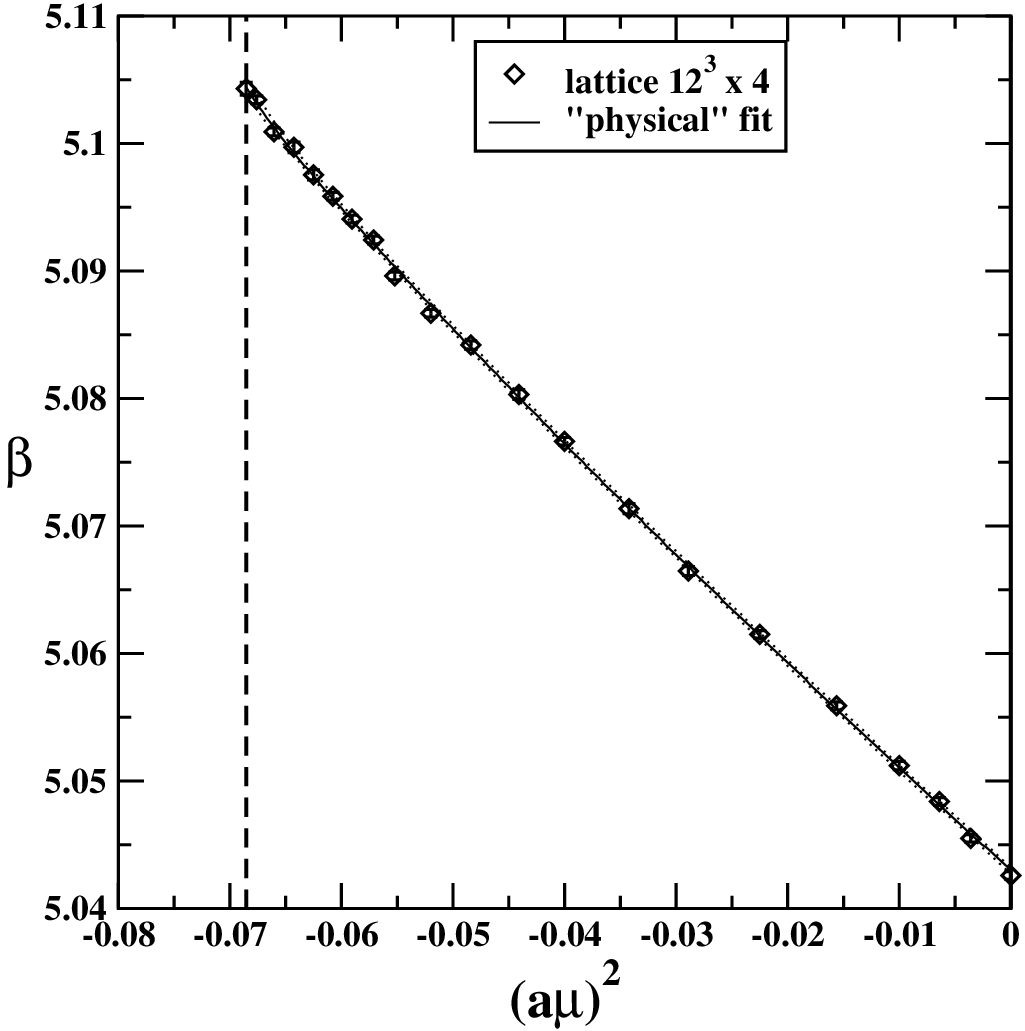}
\caption{Fits to the critical couplings: ratio of a 4th to 2th order 
polynomial (left) and ``physical'' fit according to the 
function~(\ref{phys_fit}) (right).} 
\label{fig_phys_fit}
\end{figure*}

Then, we have considered interpolations with ratios of polynomials of
order up to $\mu^{4}$ (see Table~\ref{table_fits} for a summary of the
resulting fit parameters). The interpolations with the least number
of parameters for which we got good fits to the data at $\mu^2 \leq 0$
are the ratio of a 2nd to 4th order polynomial and the ratio of a 4th to 2nd
order polynomial (see Fig.~\ref{fig_phys_fit}(left) for the latter).

Finally, we have tried here the fit strategy first suggested in 
Ref.~\cite{cea2}, consisting in writing the interpolating function in 
{\em physical units} and to deduce from it the functional dependence of 
$\beta_c$ on $\mu^2$, after establishing a suitable correspondence between 
physical and lattice units. The natural, dimensionless variables of our 
theory are $T/T_c(0)$, where $T_c(0)$ is the critical temperature at zero 
chemical potential, and $\mu/T$. The ratio $T/T_c(0)$ is deduced from 
the relation $T=1/(N_t a(\beta))$, where $N_t$ is the number of lattice sites 
in the temporal direction and $a(\beta)$ is the lattice spacing at a 
given $\beta$.
Strictly speaking 
the lattice spacing depends also on the bare quark mass, which in our runs 
slightly changes as we change 
$\beta$ since we fix $a m$. However in the following evaluation, 
which is only based on the perturbative 2-loop $\beta$-function, we shall 
neglect such dependence.
We use for $a(\beta)$ the perturbative 2-loop 
expression with $N_c=3$ and $n_f=4$.

We adopted the 3-parameter function
\beq
\left[\frac{T_c(\mu)}{T_c(0)}\right]^2=\frac{1+C\mu^2/T_c^2(\mu)}
{1+A\mu^2/T_c^2(\mu)+B\mu^4/T_c^4(\mu)}\; 
\label{phys_crit}
\eeq
leading to the following implicit relation between $\beta_c$ and 
$\mu^2$:
\begin{eqnarray}
a^2(\beta_c(\mu^2))|_{\rm 2-loop} &=& a^2(\beta_c(0))|_{\rm 2-loop}
\nonumber \\
&\times& \frac{1+A\mu^2/T_c^2 + B\mu^4/T_c^4}{1+C\mu^2/T_c^2}\;.
\label{phys_fit}
\end{eqnarray}
The values of the fit parameters turned to be
\begin{eqnarray}
\beta_c(0) &=& 5.04295(25) \;,\;\;\;\;\;\; A = 1.00315(95) \;, \nonumber \\
B &=& 0.12724(75) \;,\;\;\;\;\;\; C=0.8538(11) \;,
\label{phys_fit_param}
\end{eqnarray}
with $\chi^2$/d.o.f.=1.26. In Fig.~\ref{fig_phys_fit}(right) we compare the 
fit to data for $\beta_c(\mu^2)$. 

\begin{figure*}[htbp]
\includegraphics*[height=0.350\textheight,width=1.2\columnwidth]
{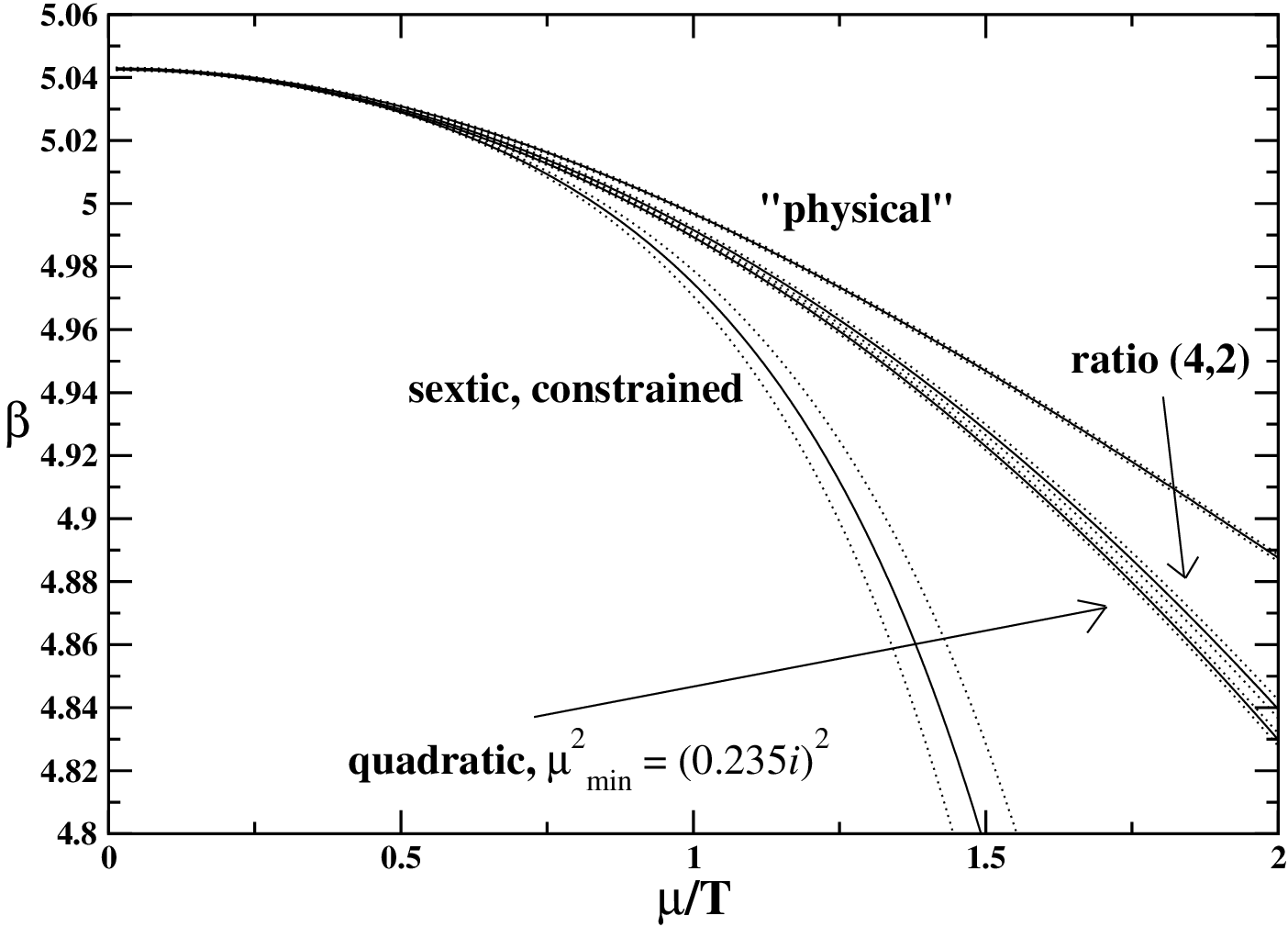}
\caption{Extrapolation to real chemical potentials of the quadratic 
(with $(a\mu)^2 = (0.235i)^2$), sextic constrained, ratio of polynomials 
and ``physical'' fits.}
\label{fig_extrapolation}
\end{figure*}

\begin{figure*}[htbp]
\includegraphics*[height=0.350\textheight,width=1.2\columnwidth]
{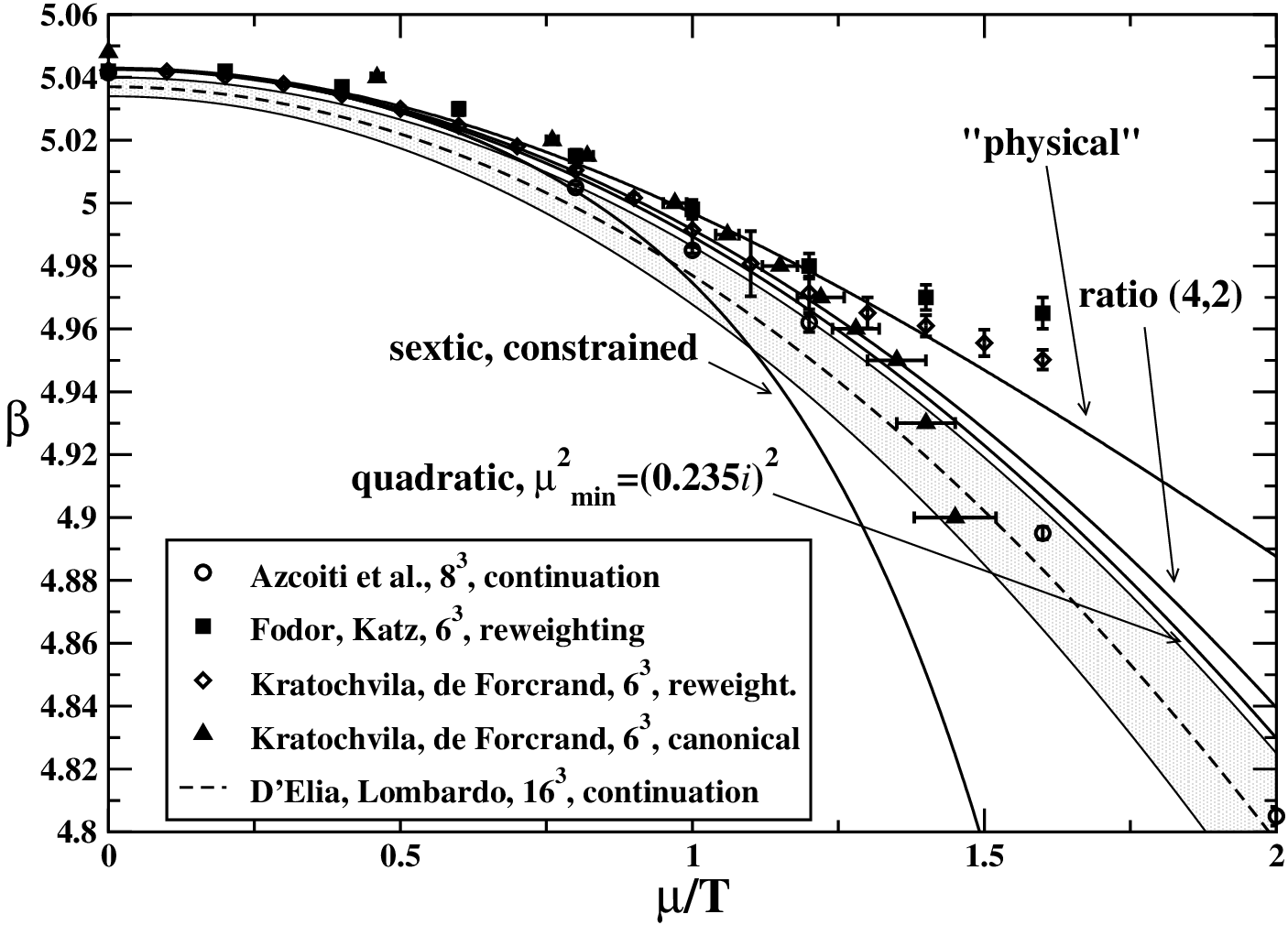}
\caption[]{Comparison of our extrapolations with other determinations in the
literature. For the sake of readability, our extrapolations have been plotted
without error bands and labels, since they can be easily recovered
from the previous figure. {\it Legenda}: 
D'Elia, Lombardo, Ref.~\cite{immu_dl}; 
Azcoiti {\it et al.}, Ref.~\cite{azcoiti}; 
Fodor, Katz, Ref.~\cite{Fodor:2001au}; 
Kratochvila, de Forcrand, Ref.~\cite{Kratochvila:2005mk}.}
\label{fig_comparison}
\end{figure*}

The important question now is whether the successful interpolations we 
found in the $\mu^2 \leq 0$ region have a consistent extrapolation to
$\mu^2 > 0$. In Fig.~\ref{fig_extrapolation} we have plotted the extrapolations
to the interval $0 \leq \mu/T \leq 2$ of the following fits: 
\begin{itemize}
\item quadratic fit, performed in the interval $-0.235^2 \leq (a\mu)^2 \leq 0$
(2nd line in Table~\ref{table_fits}); 
\item sextic constrained polynomial (5th line in Table~\ref{table_fits}); 
\item ratio (4,2) of polynomials (last line in Table~\ref{table_fits}); 
\item ``physical'' fit, Eqs.~(\ref{phys_crit})-(\ref{phys_fit_param}).
\end{itemize}
The four curves agree as long as $\mu/T \lesssim 0.6$, but
then spread. This means that different interpolations, which all reproduce
the trend of data in the fit region $-(\pi/12)^2 \leq (a\mu)^2 \leq 0$ 
and take correctly into account the deviation from the quadratic behavior
in that region, lead to somewhat distinct extrapolations. It is evident that, 
unless an extra-argument is found to make one fitting function preferable
with respect to the others, one cannot rely on a unique extrapolation, except 
in the region $\mu/T \lesssim 0.6$. The unpleasant aspect is that in the same region
also deviations from the simple quadratic behaviour are negligible: that means that,
even if we are able to see deviations from the quadratic behaviour, we are not able 
to extrapolate them to real chemical potentials in a reliable way, therefore the fact
that we can see them is in some sense useless. That emerges as a shortcoming of 
analytic continuation, which however could be less severe 
in the more physical case of $n_f = 2$ or $n_f = 2+1$. Indeed, in those cases, the 
curvature of the critical line at $\mu = 0$ (i.e. the linear term in $\mu^2$) is smaller 
than in the $n_f = 4$ case and larger non-linear contributions should be needed
to bend the critical line towards a critical baryon chemical potential
of the order of 1 GeV at $T = 0$, so that the sensitivity to such non-linear contributions
could be hopefully enhanced\footnote{We thank P.
de Forcrand for useful comments on this point, as well as for providing
us with the data needed to build up Fig.~\ref{fig_comparison}.}.

If one believes that (i) the ``physical'' fit has a better chance to give the
correct behavior of the critical line at real chemical potential and 
(ii) it reproduces the critical line all the way down to $T=0$, then 
the critical value of the chemical potential on the $T=0$ axis can be
estimated as $\sqrt{C/B}$=2.5904(93) $T_c(0)$, yielding a critical 
baryon chemical potential at $T = 0$ slightly above 1 GeV, in rough 
agreement with the expected lightest baryon mass.
Of course neither of the two assumptions above can be supported by valid arguments.

One may ask whether a different choice of the values of the imaginary chemical potential,
for which the pseudo-critical line has been located, could have improved our predictivity.
Our choice in the present work has been to distribute the values more or less uniformly
in $\mu^2$: we have then increased the density, after a preliminary analysis of our 
data, in the region closer to the RW transition, where deviations from the simple linear
deviation in $\mu^2$ were already visible. 
Of course the optimal choice could be different
from that: for instance, once the region where the linear approximation works well is known,
one realizes that fewer points chosen at the border of the same region would have provided
the same amount of information, however this is a hint which is known only {\it a 
posteriori}.

Finally, we present in Fig.~\ref{fig_comparison} an update of Fig.~4(left) of 
Ref.~\cite{Kratochvila:2005mk}, where several determinations
of the critical line existing in the literature are presented together with
the results of this work. Looking at Fig.~\ref{fig_comparison}, one could 
comment that the extrapolation of the ``physical'' fit exhibits the same trend 
as data from reweighting, whereas that from the sextic constrained fit mimics 
the strong coupling behavior~\cite{Miura:2009nu}, the other two extrapolations 
of ours lying in-between. However, previous determinations at real $\mu$ 
in the literature seem to be in fair agreement up to $\mu/T \simeq 1.2$.
If one takes this common trend as benchmark for our extrapolations,
the ``physical'' and the polynomial ratio (4,2) seem to be favoured.

We have tried to put the previous observation on a more solid ground,
including in our fit also (subsets of) data at real chemical
potential available from the literature (see Fig.~\ref{fig_comparison}).
A serious limitation of this combined approach is the inhomogeneity
of the data presently available, due to different lattices and systematics.
Indeed, we could not get acceptable values of the $\chi^2$/d.o.f.
if not restricting the region of real chemical potentials included in the fit 
to the interval $0 \leq a\mu \simeq 0.6$. Unfortunately, this is also the 
region where our extrapolations are consistent with each other, so that this 
combined fit was of little help. However, if the inhomogeneity of data at real
$\mu$ will be reduced by new Monte Carlo determinations, the combined-fit 
strategy could bring along an appreciable improvement.

\section{Discussion}

In this paper we have revisited the application of the method of analytic
continuation from imaginary to real chemical potential in QCD with $n_f=4$
degenerate flavors. The motivations of this study were
\begin{itemize}
\item to determine precisely the pseudo-critical line $\beta_c(\mu^2)$ in 
the region of negative $\mu^2$, by sampling it through the accurate 
determination by Monte Carlo methods of about 20 data points, almost 
uniformly distributed in the region $-(\pi/12)^2 \leq (a\mu)^2 \leq 0$
and by suitably interpolating these data points with an analytic function;
\item to exploit, differently from what was hitherto done in the literature,
interpolating functions sensitive to possible deviations of 
the critical line from the quadratic behavior in $\mu$ for larger absolute 
values of $\mu$ in the above-mentioned region; these deviations were clearly 
seen in QCD-like theories, such as 2-color QCD and finite isospin QCD, where 
it was given compelling evidence that their neglect could mislead the 
analytic continuation to real chemical potential; 
\item to extrapolate the newly adopted interpolations to the region of real 
$\mu$ and to re-determine, therefore, the critical line in QCD.
\end{itemize}

The outcome has been that deviations from the quadratic behavior in $\mu$
of the pseudo-critical couplings at negative $\mu^2$ are indeed visible in 
QCD with $n_f=4$. However there are several kinds of functions able to 
interpolate them, leading to extrapolations to real $\mu$ which start diverging
from each other for $\mu/T\gtrsim 0.6$. Unfortunately in the range of real chemical
potentials where the different extrapolations agree, deviations from the 
quadratic behavior in $\mu$ are negligible, so that our efforts to determine
such deviations in the critical line in a consistent way, which were successful
for QCD at finite isospin density, reveal useless in this case.

One may ask which significant differences exist between QCD at finite isospin
density and QCD at finite baryon density which could be at the basis 
of such different outcomes. Many physical properties distinguish the two
theories, but the one which is probably most relevant to our problem
is the different extension of the available region for the determination
of the critical line on the imaginary chemical potential side: for QCD
with an imaginary baryon chemical potential one has $0 < {\rm{Im}}(\mu)/T < \pi/3$, while for QCD with an imaginary isospin chemical
potential one has $0 < {\rm{Im}}(\mu)/T \lesssim \pi/2$~\cite{cea2}. In the second
case one has a wider region where more information about non-linear terms in $\mu^2$ can be
collected in order to obtain reliable extrapolations. It is indeed interesting to notice
that typical deviations of $\beta_c(\mu^2) - \beta_c(0)$ from the quadratic behaviour, visible 
at the border of the available region on the imaginary chemical potential side, are of the order 
of 5\% in the present work, while they were of the order of 10\% for QCD at finite isospin 
chemical potential.

The shortcomings of the method of analytic continuation, which emerged in
this work on QCD with $n_f=4$, could be less severe in the more physical 
case of $n_f = 2$ or $n_f = 2+1$, where the curvature of the critical line
at $\mu = 0$ (i.e. the linear term in $\mu^2$) is smaller and the sensitivity
to non-linear terms in $\mu^2$ could be enhanced.
Substantial improvement could come by either theoretical
developments, able to help discriminating between one kind of interpolation
and another, or by combined numerical strategies, aiming at gathering
information on the form of the critical line from approaches (such as 
reweighting, Taylor expansion, canonical approach, etc.) which have been 
applied so far only in exclusive way.

\section{Acknowledgments}
We are grateful to Philippe de Forcrand for useful discussions and correspondence.
We acknowledge the use of the computer facilities at the INFN apeNEXT 
Computing Center in Rome and of the PC clusters of the INFN Bari Computer 
Center for Science.

\end{document}